\documentclass[aps,prb,twocolumn,superscriptaddress,showpacs,floatfix]{revtex4}

\usepackage{graphicx}
\usepackage{epstopdf}
\usepackage{dcolumn}
\usepackage{bm}
\usepackage{float}
\usepackage{braket}

\begin{document}

\title {Electronic structure of spin frustrated magnets: Mn$_3$O$_4$ spinel and postspinel}

\author{S. Hirai}
\affiliation{Department of Geological and Environmental Sciences, Stanford University, California 94305, USA}
\affiliation{Stanford Institute for Materials and Energy Sciences, SLAC National Accelerator Laboratory, 2575 Sand Hill Road, Menlo Park, CA 94025, USA}

\author{Y. Goto}
\affiliation{Department of Applied Physics and Physico-Informatics, Keio University, 3-14-1 Hiyoshi, Yokohama 223-8522, Japan}

\author{A. Wakatsuki}
\affiliation{Department of Applied Physics and Physico-Informatics, Keio University, 3-14-1 Hiyoshi, Yokohama 223-8522, Japan}

\author{Y. Kamihara}
\affiliation{Department of Applied Physics and Physico-Informatics, Keio University, 3-14-1 Hiyoshi, Yokohama 223-8522, Japan}

\author{M. Matoba}
\affiliation{Department of Applied Physics and Physico-Informatics, Keio University, 3-14-1 Hiyoshi, Yokohama 223-8522, Japan}

\author{W. L. Mao}
\affiliation{Department of Geological and Environmental Sciences, Stanford University, California 94305, USA}
\affiliation{Stanford Institute for Materials and Energy Sciences, SLAC National Accelerator Laboratory, 2575 Sand Hill Road, Menlo Park, CA 94025, USA}

\begin{abstract}
Mn$_3$O$_4$ is a spin frustrated magnet that adopts a tetragonally distorted spinel structure at ambient conditions and a CaMn$_2$O$_4$-type postspinel structure at high pressure. We conducted both optical measurements and \emph{ab} \emph{initio} calculations, and systematically studied the electronic band structures of both the spinel and postspinel Mn$_3$O$_4$ phases. For both phases, theoretical electronic structures are consistent with the optical absorption spectra, and display characteristic band-splitting of the conduction band. The band gap obtained from the absorption spectra is 1.91(6) eV for the spinel phase, and 0.94(2) eV for the postspinel phase. Both phases are charge-transfer type insulators. The Mn 3\emph{d} $t_2$$_g$ and O 2\emph{p} form antibonding orbitals situated at the conduction band with higher energy. 

\end{abstract}


\maketitle

\section{Introduction}
\begin{figure}[!h]
\centerline{\includegraphics[scale=0.37]{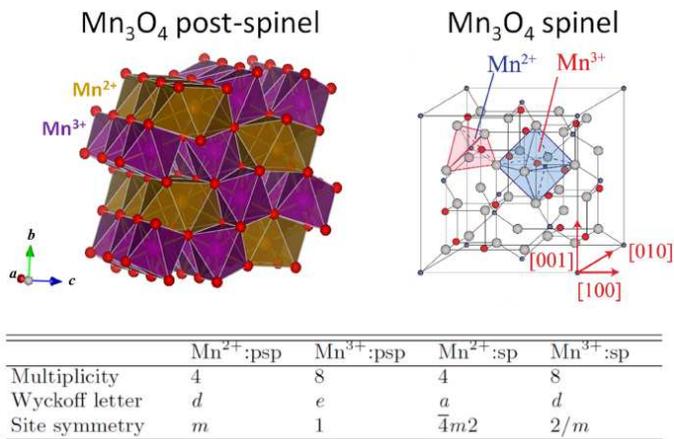}}
\caption{\label{fig:mn3o41}
(color online) Schematic image of the crystal structure of Mn$_3$O$_4$ postspinel (psp) (space group: \emph{Pbcm}(57)) and spinel (sp) (space group: \emph{I}41/\emph{amd}(141)). Multiplicity, Wyckoff letter, and site symmetry of Mn$^{2+}$ and Mn$^{3+}$ are listed as a reference.}
\end{figure}

Manganese oxides have been exploited for a wide range of technological applications including electrode materials for batteries (MnO$_2$) \cite{ammund00}, ultracapacitors for energy storage devices (MnO$_2$) \cite{ammund00}, for the catalytic combustion of methane (Mn$_2$O$_3$) \cite{ramesh07}, and as electrochromic materials (Mn$_3$O$_4$) \cite{maru95}. Mn$_3$O$_4$ is a unique mixed-valence oxide that adopts a tetragonally distorted spinel structure (Figure \ref{fig:mn3o41}) at ambient conditions \cite{amin26, xiao04, kimm11, bouch71, jensen74, char86}. This phase undergoes a structural phase transition at 15 GPa into a CaMn$_2$O$_4$-type phase, referred to as postspinel (Figure \ref{fig:mn3o41}), which is quenchable to ambient pressure \cite{paris92, merlin10, hirai13}. Lattice, spin, and orbital degrees of freedom are strongly coupled in both phases \cite{hirai13, tack07, suzuki08, niiy13, kimm10}. Mn$_3$O$_4$ spinel undergoes three magnetic transitions ($T_N = 41$ K, $T_1 = 39$ K, $T_2 = 33$ K) \cite{jensen74}, exhibits pronounced magnetodielectric and magnetoelastic coupling \cite{tack07, suzuki08, niiy13}, and a quantum phase transition \cite{kimm11, kimm10}. It adopts three magnetic structures, (i) $T_1$$<T<$$T_N$: a so-called Yafet-Kittel magnetic structure \cite{yafet52} in which Mn$^{2+}$ and Mn$^{3+}$ spins lie on the (1 -1 0) plane in a triangular spin configuration (Yafet-Kittel phase) \cite{bouch71, jensen74, char86}, (ii) $T_2$$<T<$$T_1$: a conical \cite{jensen74} or sinusoidal \cite{char86} magnetic structure incommensurate with the chemical unit cell, and (iii) $T<$$T_2$: a commensurate magnetic structure with the magnetic unit cell of the Yafet-Kittel phase doubled along the [1 1 0] direction. This complicated magnetic structure implies that Mn$_3$O$_4$ spinel is a spin frustrated magnet \cite{tack07, suzuki08, niiy13, kimm10, kuriki03, chung13}. Mn$_3$O$_4$ postspinel undergoes a giant atomic displacement near its $T_N$ = 210 K due to the coupled effect of the metastable structure with the orbital realignment of the Mn$^{3+}$ ion \cite{hirai13}. Below 210 K, the postspinel phase of Mn$_3$O$_4$ adopts a magnetic structure in which Mn$^{3+}$ spins align antiferromagnetically along the edge-sharing \emph{a} axis, with a magnetic propagation vector k =[1/2,0,0] \cite{hirai13}. In contrast, Mn$^{2+}$ spins are geometrically frustrated since they are situated at the center of a honeycomb arrangement of Mn$^{3+}$ spins that are antiferromagnetically ordered, and exhibit a short-range magnetic order only below 55 K \cite{hirai13}.

Therefore, both of these phases are magnetically frustrated \cite{hirai13, tack07, suzuki08, niiy13, kimm10, kuriki03, chung13}, providing an important playground for systematically studying the electronic band structures of spin frustrated magnets \cite{nisoli13}. Electrical resistivity for the bulk polycrystalline samples are $\approx10^8$ $\Omega$cm \cite{rozh66} for the spinel phase, and $\approx10^7$ $\Omega$cm \cite{hirai13} for the postspinel phase. Previous studies on the band structure of Mn$_3$O$_4$ spinel are limited to theoretical studies using Hartree-Fock \cite{chart99} and density functional theory (DFT) calculations \cite{franch07}, which have significant disagreement in their results.  While there are a few studies reported for thin film ($E_g = 2.51$ eV) \cite{dubal10} and nanoparticles ($E_g = 2.07$ eV) \cite{jhaa12} for Mn$_3$O$_4$ spinel, no optical studies have been reported for bulk polycrystalline and single crystalline samples of Mn$_3$O$_4$ spinel or postspinel. 
      
In this study, we found that both phases of Mn$_3$O$_4$ possess characteristic band-splitting of the conduction band. Theoretical electronic structures are consistent with absorption spectra, where the band gap is 1.91(6) eV for the spinel phase and 0.94(2) eV for the postspinel phase. Both phases are charge-transfer type insulators. Mn 3\emph{d} $t_2$$_g$ and O 2\emph{p} form antibonding orbitals situated at the conduction band with higher energy.

\section{Experiments}
Polycrystalline samples of Mn$_3$O$_4$ spinel and postspinel were prepared at ambient and high pressure, respectively \cite{hirai13}. The spinel phase was prepared by heating MnCO$_3$ (Alfa Aesar, 99.99 \%) in air at 1400 K for 16 h. The postspinel phase was prepared by pressurizing the spinel phase up to 20 GPa using a cell fitted with polycrystalline diamond double toroidal anvils, followed by slow decompression to ambient pressure. Purity of the samples was examined by using synchrotron X-ray diffraction (XRD) ($\lambda = 0.041222$ nm) at beamline 16-BMD of the advanced photon source (APS), Argonne National Laboratory (ANL) \cite{hirai13}. 

Total diffuse reflectance spectra of Mn$_3$O$_4$ spinel and postspinel were measured using a spectrometer equipped with an integrating sphere (Hitachi High-Tech, U-4100). Al$_2$O$_3$ powder was used as a standard reference. Then, the diffuse reflectance spectra were converted to absorption spectra using the Kubelka-Munk equation, (1-\emph{R})$^2$/2\emph{R} = $\alpha$/\emph{s}, where \emph{R}, $\alpha$ and \emph{s} denote reflectivity, absorption coefficient and scattering factor, respectively \cite{kubelk31}. The optical band gap was deduced from the linear extrapolation of absorption spectra with energy.

\emph{Ab} \emph{initio} calculations were performed using the plane-wave projector augmented-wave (PAW) method \cite{blochl94, kresse99} implemented in Vienna \emph{Ab} \emph{initio} Simulation Package (VASP) code \cite{kresse96, kress96}. Three different approximations to treat exchange-correlation potential were applied: (i) the generalized gradient approximation (GGA) by the Perdew-Becke-Ernzerhof (PBE) method \cite{perdew96}, (ii) the PBE plus a on-site Coulomb repulsion (U) term following the approach of Dudarev \cite{dudar98} for effective value of the Coulomb parameter ($U_{eff}$ = U-J, where J being an intra-atomic exchange parameter), and (iii) the hybrid functionals by the Heyd, Scuseria, and Ernzerhof (HSE06) method \cite{heyd03}, where correlation is described in GGA (PBE) and its exchange is a mixture of 25 \% exact Hartree-Fock and 75 \% PBE exchange. We used the lattice constants of \emph{a} = 0.5756 nm and \emph{c} = 0.9439 nm at 4.2 K \cite{bouch71} for the spinel phase and \emph{a} = 0.3020 nm, \emph{b} = 0.9842 nm, and \emph{c} = 0.9568 nm at 60 K \cite{hirai13} for the postspinel phase, which were determined by neutron diffraction. We used the lattice constants at 60 K for the postspinel phase since the atomic coordinates could not be determined from neutron diffraction due to the contribution of short-range ordered Mn$^{2+}$ spins \cite{hirai13}. Also a 2$\times{1}$$\times{1}$ supercell was constructed to investigate the magnetic structure of postspinel. The Brillouin zone was sampled by a 3$\times{3}$$\times{2}$ and a 3$\times{3}$$\times{2}$ Monkhorst-Pack grid \cite{monkh76} for the spinel and postspinel phases, respectively. A cutoff of 450 eV was chosen for the plane-wave basis set for both phases. Hellmann-Feynman forces were reduced until 0.5 eV nm$^{-1}$ for the spinel phase, while the atomic coordinates for the postspinel phase were fixed at the initial input value due to its metastable nature \cite{hirai13}.

\section{Results and discussion}
\begin{figure}[!h]
\centerline{\includegraphics[scale=0.35]{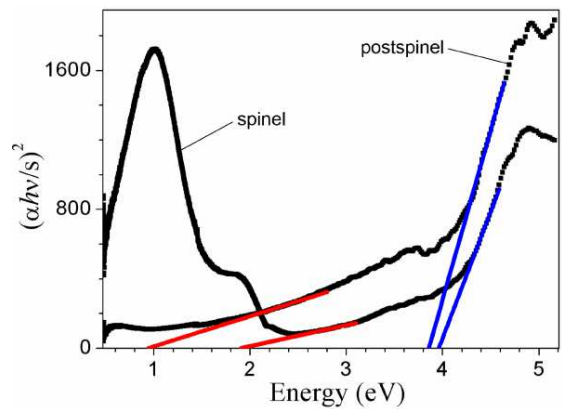}}
\caption{\label{fig:mn3o42}
(color online) Absorption spectra of Mn$_3$O$_4$ spinel and postspinel, converted from diffuse reflectance spectra using the Kubelka-Munk equation, (1-\emph{R})$^2$/2\emph{R} = $\alpha$/\emph{s}, where \emph{R}, $\alpha$ and \emph{s} denote reflectivity, absorption coefficient and scattering factor, respectively \cite{kubelk31}. Red and blue lines are extrapolated to the energy axis, providing an estimate of optical band gap for these phases.}
\end{figure}

\begin{figure}[!h]
\centerline{\includegraphics[scale=0.65]{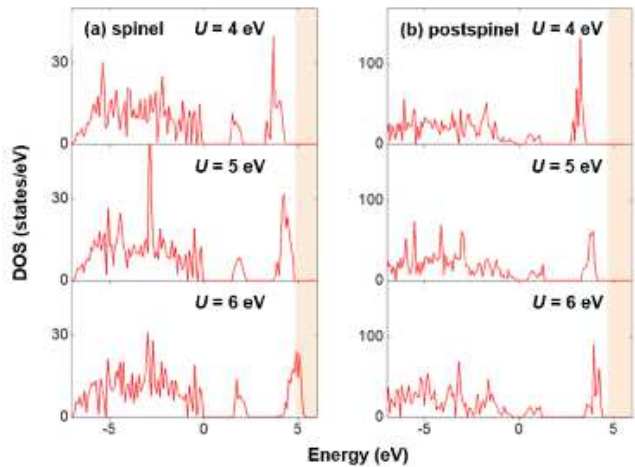}}
\caption{\label{fig:mn3o43}
(color online) Up-spin densities of states in Mn$_3$O$_4$ spinel and postspinel calculated with U-J = 4, 5, 6 eV in PBE+U method. The work function of manganese oxides (e.g. MnO) is in a range of 4.6-6.6 eV \cite{toroker11}, shaded pink as a reference.}
\end{figure}

Figure \ref{fig:mn3o42} shows the absorption spectra of Mn$_3$O$_4$ spinel and postspinel, converted from diffuse reflectance spectra using the Kubelka-Munk equation \cite{kubelk31}. Red and blue lines are guide for estimating the optical band gap of the two energy levels of the conduction band in these phases. \emph{Ab} \emph{initio} calculations were also performed in order to obtain the theoretical band gap. Table \ref{tab:paramtab} shows the experimental band gap from optical absorption spectra and the theoretical band gap from \emph{ab} \emph{initio} calculations for Mn$_3$O$_4$ spinel and postspinel. Both methods show a consistent band structure with characteristic band-splitting of the conduction band (Table \ref{tab:paramtab} and Figure \ref{fig:mn3o43}). The band gap obtained from optical absorption spectra is 1.91(6) eV for the spinel phase and 0.94(2) eV for the postspinel phase. The calculated band structure will be presented for both phases, and then their electronic structures will be further discussed.

\begin{table*}[tp]
\centering
\begin{tabular}{l@{\hspace{1cm}}       l@{\hspace{1cm}}           l@{\hspace{1cm}}           l@{\hspace{1cm}}        l@{\hspace{1cm}}}
  \hline\hline
                Methods  & $E_{g1}$:sp (eV)  & $E_{g2}$:sp (eV)  &$E_{g1}$:psp (eV)  &$E_{g2}$:psp (eV)    \\
  \hline
                 U-J=4 eV   & 1.36  & 3.17  & 0.36  & 2.81  \\
                 U-J=5 eV   & 1.46  & 3.47  & 0.55  & 3.40 \\
								 U-J=6 eV   & 1.48  & 3.60  & 0.56  & 3.71  \\
								 HSE06   & 2.61  & 4.17  & 1.34  & 3.56  \\
								 Experimental   & 1.91(6)  & 3.96(9)  & 0.94(2)  & 3.86(6)  \\
								
  \hline\hline
\end{tabular}
\caption{Theoretical (calculated by PBE+U, HSE06) and experimental (deduced from optical absorption spectra) band gap for Mn$_3$O$_4$ spinel (sp) and postspinel (psp). The optical band gaps for the two energy levels of the conduction band are denoted as $E_{g1}$ and $E_{g2}$, respectively.}
\label{tab:paramtab}
\end{table*}

\begin{figure}[!h]
\centerline{\includegraphics[scale=0.7]{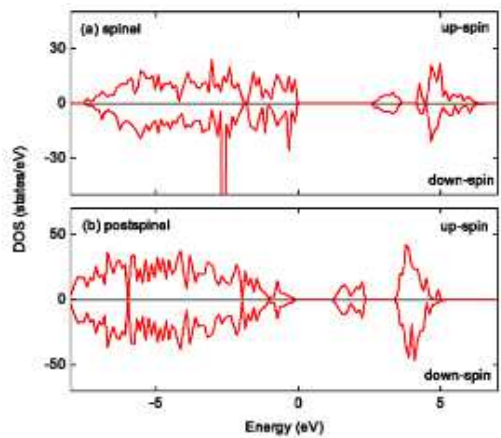}}
\caption{\label{fig:mn3o43s}
(color online) Densities of states in Mn$_3$O$_4$ spinel and postspinel calculated in HSE06 method.}
\end{figure}

\begin{figure}[!h]
\centerline{\includegraphics[scale=0.46]{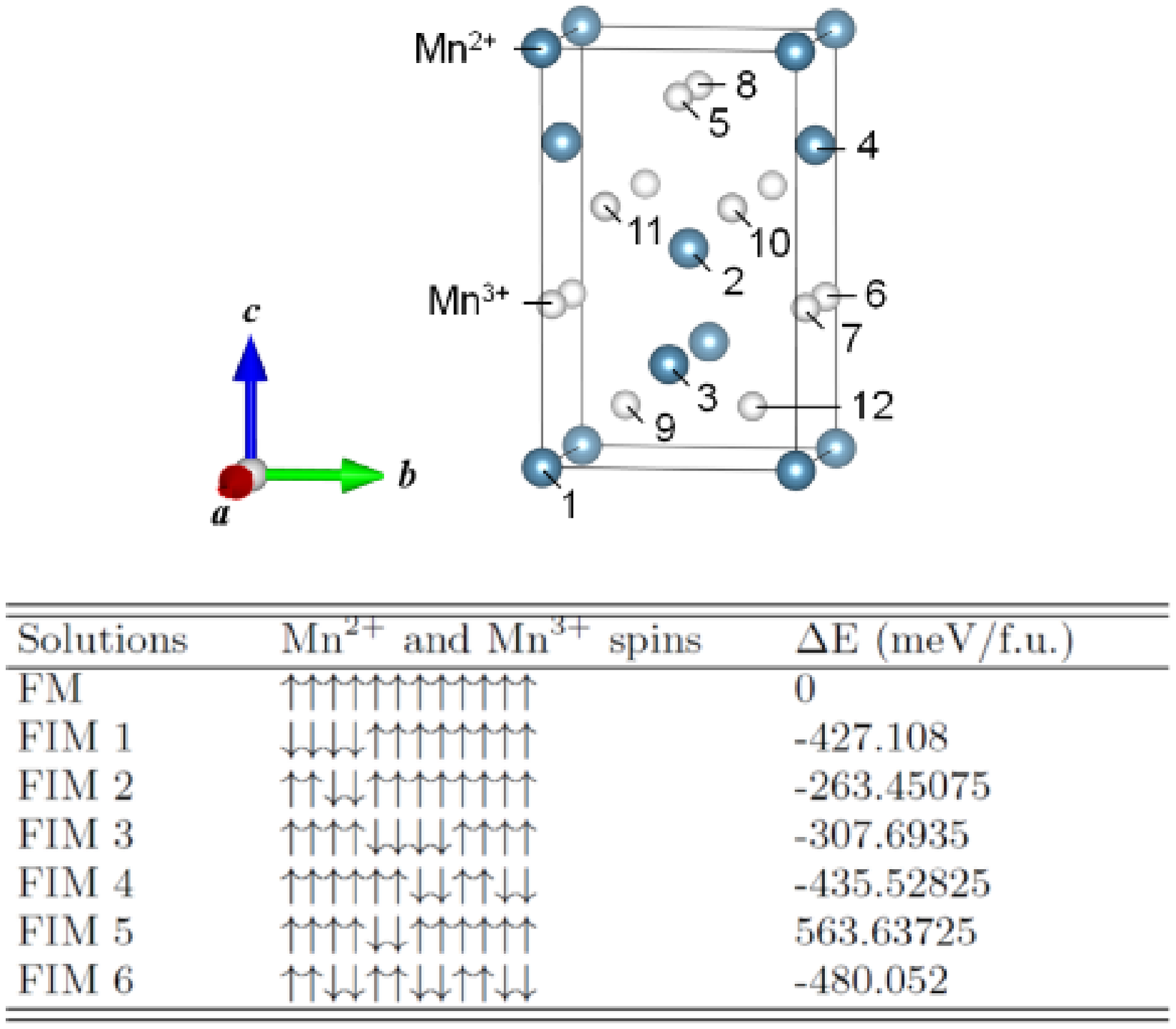}}
\caption{\label{fig:mn3o44}
(color online) Spin configurations of Mn and the total energy differences ($\Delta$E) with respect to the ferromagnetic (FM) solutions for Mn$_3$O$_4$ spinel with U-J = 5 eV in PBE+U method. The calculated magnetic moment of Mn$^{2+}$ and Mn$^{3+}$ are 4.6 $\mu_B$/atom and 3.9 $\mu_B$/atom, respectively. The calculated magnetic moment of Mn$^{2+}$ and Mn$^{3+}$ are in good agreement with the previously reported ordered moment of 4.65 $\mu_B$/atom and 3.55 $\mu_B$/atom at 4.2 K obtained from neutron diffraction \cite{bouch71}.}
\end{figure}

\begin{figure}[!h]
\centerline{\includegraphics[scale=0.46]{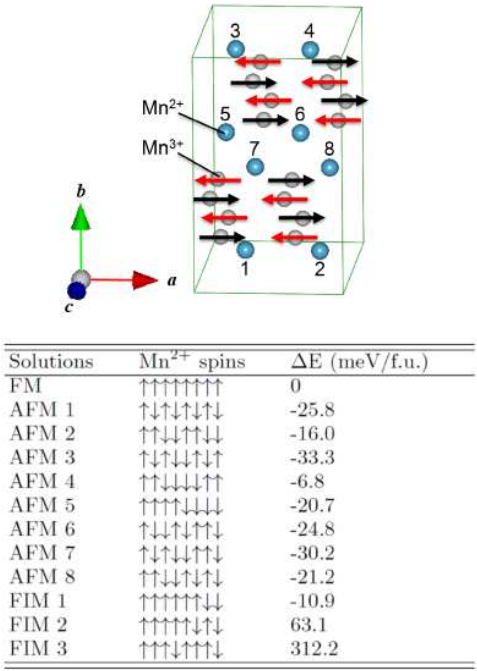}}
\caption{\label{fig:mn3o45}
(color online) Spin configurations of Mn$^{2+}$ and the total energy differences ($\Delta$E) with respect to the ferromagnetic (FM) solutions for Mn$_3$O$_4$ postspinel with U-J = 5 eV in PBE+U method. Black and red arrows denote the spin configuration of Mn$^{3+}$ , which was fixed to an antiferromagnetic (AFM) order along the crystallographic a axis, with a magnetic propagation vector k =[1/2,0,0] \cite{hirai13}. The calculated magnetic moment of Mn$^{2+}$ and Mn$^{3+}$ are 4.7 $\mu_B$/atom and 4.0 $\mu_B$/atom, respectively. The calculated magnetic moment of Mn$^{3+}$ is in good agreement with the previously reported ordered moment of 3.49(2) $\mu_B$/atom at 60 K obtained from neutron diffraction \cite{hirai13}. The calculated magnetic moment of Mn$^{2+}$ does not contradict with the short-range magnetic order of Mn$^{2+}$ spins below 55 K \cite{hirai13} since the spin canting was not considered in PBE method.}
\end{figure}

\begin{figure}[!h]
\centerline{\includegraphics[scale=0.6]{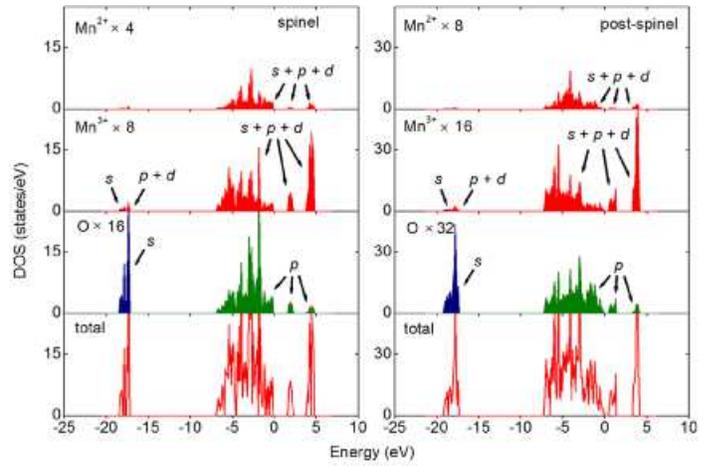}}
\caption{\label{fig:mn3o46}
(color online) Up-spin DOS calculated with U-J = 5 eV for each atom (Mn$^{2+}$, Mn$^{3+}$ and O) and total DOS in Mn$_3$O$_4$ spinel and postspinel. More than 80 \% of the \emph{s}+\emph{p}+\emph{d} orbitals of Mn are 3\emph{d} orbitals.}
\end{figure}

\begin{figure}[!h]
\centerline{\includegraphics[scale=0.5]{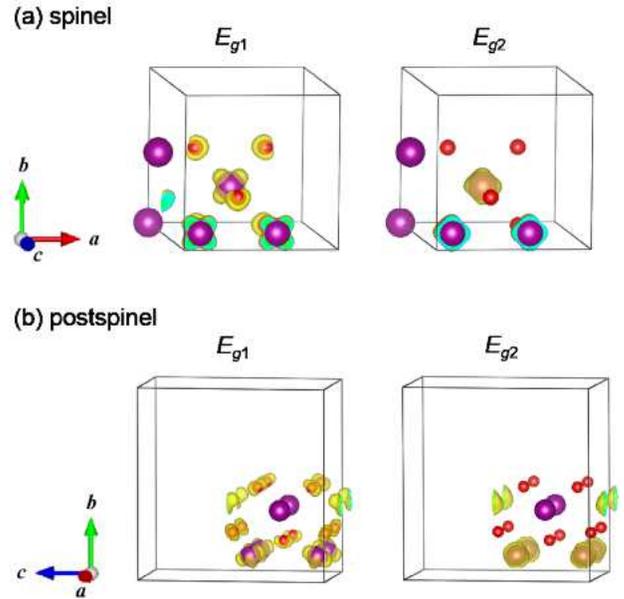}}
\caption{\label{fig:mn3o47}
(color online) Theoretical local charge density of Mn$^{3+}$O$_6$ octahedra in Mn$_3$O$_4$ spinel and postspinel integrated over the two energy levels of the conduction band, denoted as $E_{g1}$ and $E_{g2}$. Yellow shading represents the charge distribution in Mn and O.}
\end{figure}

\begin{figure}[!h]
\centerline{\includegraphics[scale=0.5]{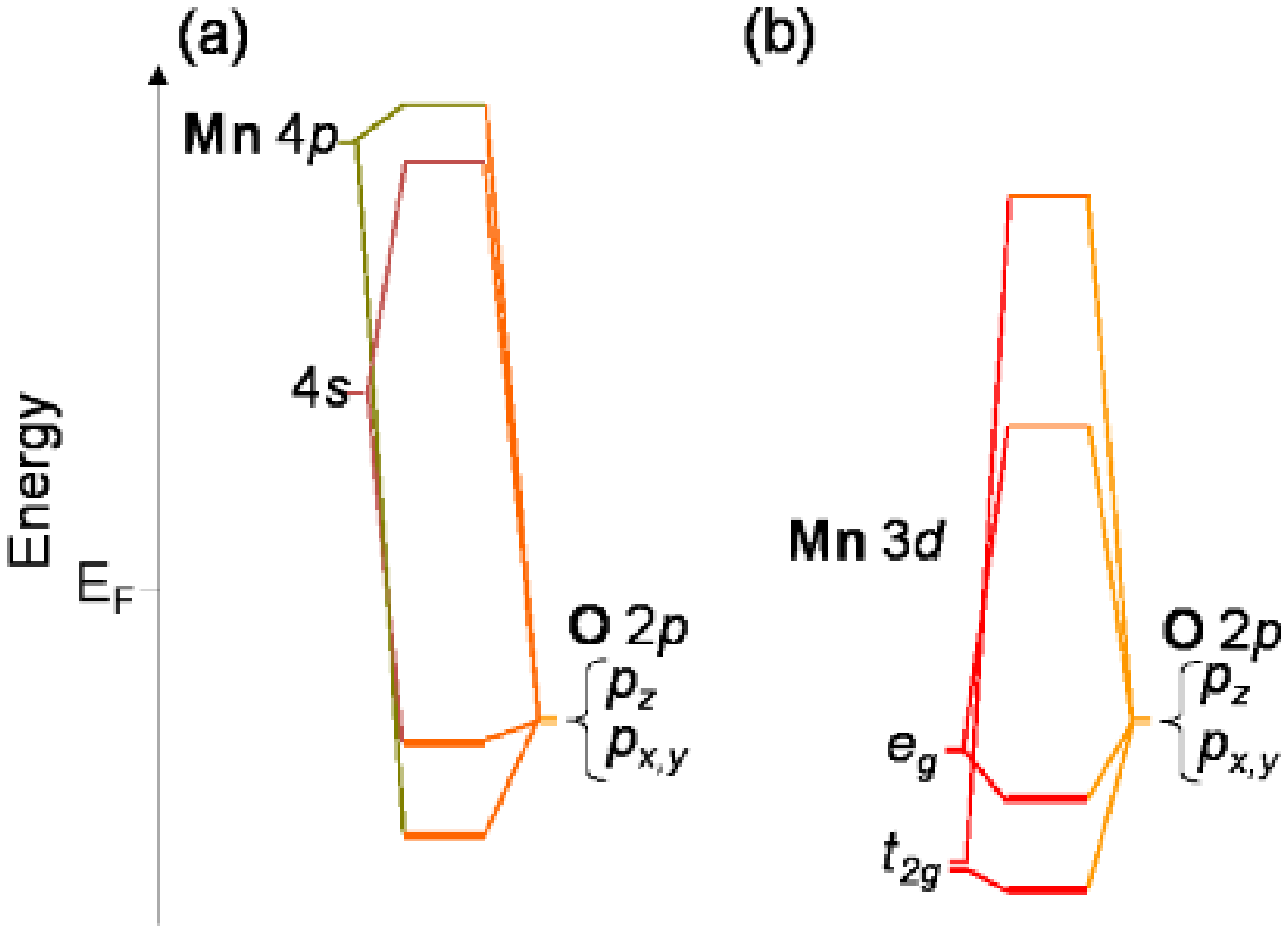}}
\caption{\label{fig:mn3o48}
(color online) Schematic view of the energy diagram of Mn and O for both phases in Mn$_3$O$_4$, provided that Mn 3\emph{d} orbitals pointing toward the O atom are $t_2$$_g$ and the rest are $e_g$. (a) An extracted diagram for Mn 4\emph{s}, Mn 4\emph{p}, and O 2\emph{p}. (b) An extracted diagram for Mn 3\emph{d} and O 2\emph{p}. }
\end{figure}

First, the band structure of Mn$_3$O$_4$ spinel will be discussed. Figure \ref{fig:mn3o43}(a) shows the DOS calculated for Mn$_3$O$_4$ spinel with U-J = 4, 5, 6 eV. HSE06 calculations (Figure \ref{fig:mn3o43s}(a)) was conducted as well as PBE+U calculations, and all these calculations give a widely spread valence band and a characteristic band-splitting of the conduction band (Figure \ref{fig:mn3o43}(a) and Figure \ref{fig:mn3o43s}(a)). This band-splitting of the conduction band is allowed since the work function of manganese oxides (e.g. MnO) is in a range of 4.6-6.6 eV \cite{toroker11}. The band gap and the energetically most stable magnetic structure (Figure \ref{fig:mn3o44}, referred to as FIM 6 in a previous study \cite{chart99}) of Mn$_3$O$_4$ spinel obtained from the \emph{ab} \emph{initio} calculations is consistent with Chartier et al \cite{chart99} and the previously reported cell-doubled magnetic structure below 33 K obtained from neutron diffraction studies \cite{bouch71, jensen74, char86}. 

Next, the band structure of Mn$_3$O$_4$ postspinel will be discussed. The band gap of the postspinel phase (0.94(2) eV from optical absorption spectra) is about 1 eV smaller than that of the spinel phase (Table \ref{tab:paramtab}). Figure \ref{fig:mn3o43}(b) and Figure \ref{fig:mn3o43s}(b) show the DOS calculated for Mn$_3$O$_4$ postspinel, where all the calculations give a wide valence band and characteristic band-splitting of the conduction band, similar to what was observed for the spinel phase. The two most stable magnetic structure calculated by this study (AFM 3 and AFM 7 in Figure \ref{fig:mn3o45}) does not contradict with the short-range magnetic order of Mn$^{2+}$ spins below 55 K since the canted Mn$^{2+}$ spins allow ferromagnetic net magnetization \cite{hirai13}. The energy difference between AFM 3 and AFM 7 (Figure \ref{fig:mn3o45}) is only 3.1 meV/f.u., which is comparable with the thermal fluctuation of $\approx{4.7}$ meV at 55 K. Therefore, \emph{ab} \emph{initio} calculation demonstrates that Mn$_3$O$_4$ postspinel is a spin frustrated magnet and the small energy difference between two AFM solutions is in harmony with the experimentally observed short-range magnetic order below 55 K \cite{hirai13}. Since the calculated band gap is in good agreement with the optical absorption spectra and the most stable magnetic structure does not contradict with the neutron diffraction studies for both phases \cite{bouch71, jensen74, char86, hirai13}, the electronic band structure of Mn$_3$O$_4$ is further explored on the basis of \emph{ab} \emph{initio} calculation. Hereafter we will employ the up-spin DOS with U-J = 5 eV for the phases, since the selection of U-J parameter does not change its DOS structure to a great extent. Figure \ref{fig:mn3o46} demonstrates that Mn 3\emph{d} orbitals are widely spread over the valence band, and they have significant overlap with the O 2\emph{p} orbitals. Although the band width becomes narrower, this feature is maintained for the conduction band as well. 

Mn$_3$O$_4$ spinel has a tetragonal symmetry (space group: \emph{I}41/\emph{amd}(141)), which results in the Mn 3\emph{d} orbitals being split into three nondegenerate levels and a doubly degenerate level. Figure \ref{fig:mn3o47}(a) shows the theoretical local charge density map of Mn$^{3+}$O$_6$ octahedra calculated for the two energy levels of the conduction band in Mn$_3$O$_4$ spinel. Since the theoretical DOS of Mn$^{2+}$ electrons in the conduction band is very small, the theoretical DOS of Mn$^{3+}$  electrons sufficiently represent the total DOS for Mn 3\emph{d} electrons in the conduction band. The multiplicity, Wyckoff letter, and site symmetry for Mn$^{3+}$  in Mn$_3$O$_4$ spinel are 8, \emph{d}, and 2/\emph{m}, respectively. Provided that the Mn 3\emph{d} orbitals pointing toward the O atom are $t_2$$_g$ and the rest are $e_g$, the projection of charge distribution for Mn 3\emph{a} electrons (Figure \ref{fig:mn3o47}(a)) suggests that the conduction band with lower energy ($E_{g1}$) corresponds to Mn 3\emph{d} $t_2$$_g$, and the conduction band with higher energy ($E_{g2}$) corresponds to Mn 3\emph{d} $e_g$ (Figure \ref{fig:mn3o48}). 

Mn$_3$O$_4$ postspinel has an orthorhombic symmetry (space group: \emph{Pbcm}(57)), which results in the Mn 3\emph{d} levels being five nondegenerate levels. The multiplicity, Wyckoff letter, and site symmetry of Mn$^{3+}$  in Mn$_3$O$_4$ postspinel are 8, \emph{e}, and 1, respectively. Analogous to the spinel phase, the projection of charge distribution for Mn 3\emph{d} electrons of Mn$_3$O$_4$ postspinel (Figure \ref{fig:mn3o47}(b)) suggests that the conduction band with lower energy ($E_{g1}$) corresponds to Mn 3\emph{d} $t_2$$_g$, and the conduction band with higher energy ($E_{g2}$) corresponds to Mn 3\emph{d} $e_g$ (Figure \ref{fig:mn3o48}). Therefore, for both phases, Mn 3\emph{d} orbitals overlap with O 2\emph{p} orbitals in the valence band at the lower energy level of $t_2$$_g$ and $e_g$. In other words, the Mn 3\emph{d} levels split into bonding and antibonding levels forming hybridized orbitals with ligand O 2\emph{p} orbitals. Both phases of Mn$_3$O$_4$ are charge-transfer type insulators with characteristic band-splitting of the conduction band (Figure \ref{fig:mn3o43} and Figure \ref{fig:mn3o46}). 

\section{Conclusion}
Electronic structures of Mn$_3$O$_4$ spinel and postspinel were systematically studied. \emph{Ab} \emph{initio} calculations and the optical measurements were performed on both phases, resulting in a consistent band structure with characteristic band-splitting of the conduction band. We obtained the band gap of 1.91(6) eV for the spinel phase, and 0.94(2) eV for the postspinel phase. Both phases of Mn$_3$O$_4$ are charge-transfer type insulators, and Mn 3\emph{d} $t_2$$_g$ and O 2\emph{p} form antibonding orbitals situated at the conduction band with higher energy. \emph{Ab} \emph{initio} calculations also demonstrate that Mn$_3$O$_4$ postspinel is a spin frustrated magnet. 

\section{Acknowledgements}
This research is funded by the U.S. Department of Energy (DOE), Office of Basic Energy Sciences (BES). S.H. and W.L.M. are supported by the U.S. Department of Energy (DOE), Office of Basic Energy Sciences (BES), Division of Materials Sciences and Engineering, under Contact No. DE-AC02-76SF00515.

\end{document}